\RequirePackage{fix-cm}
\documentclass[smallextended]{svjour3}       
\smartqed  
\usepackage{times}
\usepackage{graphicx}
\usepackage{epsf}
\usepackage[colorlinks={true}]{hyperref}

\usepackage[T1]{fontenc}
\usepackage[utf8]{inputenc}
\hypersetup{citecolor={blue}, filecolor={blue}, linkcolor={blue}, urlcolor={blue}}

\newcommand{\commentold}[1]{}
\DeclareMathSymbol{:}{\mathpunct}{operators}{"3A}
\begin{document}

\title{Local quantum Fisher information and local quantum uncertainty in
two-qubit Heisenberg XYZ chain with
Dzyaloshinskii–Moriya interactions
}


\author{Soroush  Haseli$^{1}$   
       }


\institute{S.  Haseli$^{1}$ \at
              $^{1}$Faculty of Physics, Urmia University of Technology, Urmia, Iran. \\
              \email{soroush.haseli@uut.ac.ir}           
}

\date{Received: date / Accepted: date}

\maketitle

\begin{abstract}
 In this work we will study the Local quantum Fisher information and the local quantum uncertainty in two-qubit Heisenberg $XYZ$ spin chain model with Dzyaloshinskii–Moriya (DM) interaction in $z$-direction. Here we show that the $DM$ interaction and spin interactions along the $x$, $y$, and $z$ axis can increase and maintain local quantum Fisher information and local quantum uncertainty. It is also shown that because of thermal fluctuations local quantum Fisher information and local quantum uncertainty are decreased by increasing temperature. They are equal to one at very low temperature and starts to decay only after a thershold temperature.
\end{abstract}
\section{Introduction}	
Quantum entanglement is a special type of quantum correlation that plays an important role in quantum information theory. Quantum entanglement has a wide range of applications in quantum information tasks, such as quantum computing \cite{Nielsen}, quantum communications \cite{Braunstein,Bouwmeester} and and quantum key distribution \cite{Ekert}. Much recent works have focused on investigating and quantifying quantum entanglement for multipartite closed and open quantum systems \cite{Yu,Wootters,Viola,Carvalho}. Recent studies also showed that quantum entanglement is not the only quantum correlation in quantum information theory\cite{Datta,Lanyon,Datta1,Ollivier}. It has been shown that there are some separable quantum states that have quantum correlation despite the lack of quantum entanglement \cite{Lanyon,Datta1}. So, introducing a suitable criterion for determining quantum correlations beyond entanglement  has been the subject of many efforts. In Ref. \cite{Ollivier} Ollivier and Zurek, have introduced Quantum discord as the quantum correlation beyond entanglement. In Ref.\cite{Streltsov}, it is shown that classical means can not communicate the measurement results completely if a measurement apparatus  is in a nonclassical state. This means that information loss occurs even when the measuring device is not entangled  with the system, and this lost information is the quantum discord. It has also been shown that quantum discord is  more robust than quantum entanglement in dissipative systems\cite{Werlang}. Entropic quantum discord is introduced as the difference between quantum mutual information and classical information. Obtaining an analytical expression for quantum discord is only possible for certain class of states, and the situation is somewhat complicated for general state.  Difficulty in calculating quantum discord is due to optimization process over all local generalized measurements. This difficulty in calculating led to an alternative definition for quantum discord that is called geometric measure of quantum discord \cite{Huang}. It is introduced as the minimum  distance between the given state and the zero discord state. Analytically, calculating this geometric criterion requires a simpler optimization process than entropic quantum discord. Despite this advantage of geometric quantum discord, this criterion cannot be a suitable criterion for showing non-classical correlations\cite{Piani}. 

To address these difficulties, some methods and tools have been proposed to identify non-classical correlations. In Ref. \cite{Girolami}, the authors have introduced the notion of local quantum uncertainty (LQU) as a discord-like measure of non-classical correlation.  It is defined as the minimum uncertainty induced by applying local measurements on one part of quantum state using the concept of Wigner-Yanase skew information \cite{Wigner}. This measure meets all the conditions required for a measure of quantum correlations. In addition, LQU is associated to quantum Fisher information (QFI). It has been shown that in the unitary evolution of the density matrix i.e. $\rho_\theta = e^{-iH \theta}\rho e^{i H \theta}$, the QFI associated with the phase parameter majorizes the skew information. In Ref. \cite{Kim}, it has been shown that local quantum Fisher information (LQFI) can be used to describe quantum correlations based on QFI. This measure is defined based on the optimizations over the observables related to one of the subsystems. In addition, local quantum Fisher information provides a tool for understanding the role of quantum correlations beyond entanglement in improving the accuracy and efficiency of quantum metrology protocols. Given that quantum Fisher's information and local quantum uncertainty are both based on the concept of quantum uncertainty and the quantify non-classical correlations, it is important to study these concepts in multipartite quantum systems. In Ref. \cite{Slaoui}, the authors study the LQU and LQFI in in Heisenberg XYmodel. They showed that LQU and LQFI depend on the temperature and the coupling parameter in the anisotropic XY model. They showed that for high temperatures, the quantum correlation decreases and reaches zero.

In this work we will study the local quantum uncertainty and local quantum Fisher information in two-qubit Heisenberg XYZ chain with Dzyaloshinskii–Moriya interactions. The work is organized as follow. In Sec. \ref{Sec2}, In Section 2, we will review the concepts of LQFI and LQU as the measures of quantum correlations. In Sec.\ref{Sec3}, we introduce the
two-qubit Heisenberg $XYZ$ spin system with DM interaction alone  $z$-direction. We also study the LQFI and LQU for two-qubit Heisenberg $XYZ$ spin system with DM interaction in this section. In Sec.
\ref{Sec4}, we summarize the results.
\section{Quantum uncertainty and quantum correlation}\label{Sec2}
\subsection{Local quantum Fisher information}
Quantum Fisher information is a practical quantity for describing optimal accuracy in parameter estimation protocols \cite{Helstrom,Kay,Genoni}. Many attempts have been made to investigate the evolution of QFI to determine the relationship between quantum entanglement and quantum metrology \cite{Chapeau1,Giovannetti}. In Refs. \cite{Huelga,Chapeau}, it has been shown that in the unitary evolution, quantum entanglement leads to a significant improvement in efficiency and accuracy of parameter estimation. For a desired parametric state $\rho_\theta$ that depends on $\theta$, the QFI is defined as follows
\begin{equation}
\mathcal{F}^{2}(\rho_\theta)=\frac{1}{4}tr(\rho_\theta L_\theta^{2}),
\end{equation}
where $L_\theta$ is the symmetric logarithmic derivative operator. $L_\theta$ is characterized as the solution of the equation 
\begin{equation}
\frac{\partial \rho_\theta}{\partial \theta} =\frac{1}{2}(L_\theta \rho_\theta + \rho_\theta L_\theta).
\end{equation} 
The parametric state $\rho_\theta$ can be obtain by the effect of the unitary evolution $U_\theta=e^{i H \theta}$ on $\rho$ as $\rho_\theta=U_\theta^{\dag}\rho U_\theta$. For a given quantum state $\rho=\sum_m \lambda_m \vert m \rangle \langle m \vert$ with $\lambda_m \geq 0$ and $\sum_m \lambda_m=1$, the QFI $\mathcal{F}^{2}(\rho_\theta)$, that we denote by $\mathcal{F}^{2}(\rho, H)$, is obtained as 
\begin{equation}
\mathcal{F}^{2}(\rho, H)=\frac{1}{2} \sum_{m \neq n} \frac{\left(\lambda_{m}-\lambda_{n}\right)^{2}}{\lambda_{m}+\lambda_{n}}|\langle m|H| n\rangle|^{2}.
\end{equation}
Let us consider the $M \times N$ bipartite quantum state $\rho=\sum_m \lambda_m \vert \lambda_m \rangle \langle \lambda_m \vert$ in the Hilbert space $\mathcal{H}_A^{M} \otimes \mathcal{H}_B^{N}$. We supposed that the dynamics of first part is described with the Hamiltonian $H_A=H_A \otimes I_B$. In this case the local quantum Fisher information (LQFI) can be written as \cite{Bera}. 
\begin{equation}\label{LQFI}
\mathcal{F}^{2}(\rho, H_{A})=tr\left(\rho H_{A}^{2}\right)-\sum_{m \neq n} \frac{2 \lambda_{m} \lambda_{n}}{\lambda_{m}+\lambda_{n}}\left|\left\langle m\left|H_{A}\right| n\right\rangle\right|^{2},
\end{equation}
LQFI is used to Characterize non-classical correlations\cite{Kim}. This quantity has special properties that any suitable correlation quantifier must have these properties. It is possible to define a quantum correlation quantifier based on LQFI by minimizing LQFI over all local Hamiltonians $H_A$
\begin{equation}
\mathcal{Q}_A^{2}=\min_{H_A}\mathcal{F}^{2}(\rho,H_A),
\end{equation}
$\mathcal{Q}_A^{2}$ vanishes for classical-quantum and classical-classical states\cite{Bera}. $\mathcal{Q}_A^{2}$ can be obtained easily for a bipartite quantum state with $2 \times N$ dimension. The overall shape of the local Hamilton is $H_A=\vec{\sigma}.\vec{r}$, where $\vert \vec{r} \vert=1$ and $\vec{\sigma}=(\sigma_1,\sigma_2,\sigma_3)$ are usual Pauli matrices. It can be seen that the first term in Eq.(\ref{LQFI}) is equal to one, i.e. $tr\left(\rho H_{A}^{2}\right)=1$ and the second term is
\begin{eqnarray}
\sum_{m \neq n} \frac{2 \lambda_{m} \lambda_{n}}{\lambda_{m}+\lambda_{n}}\left|\left\langle m\left|H_{A}\right| n\right\rangle\right|^{2}&=& \\
&=&\sum_{i,j=1}^{3}\sum_{m \neq n}\frac{2 \lambda_{m} \lambda_{n}}{\lambda_{m}+\lambda_{n}}\left\langle m\left|\sigma_{i} \otimes I\right| n\right\rangle\left\langle n\left|\sigma_{j} \otimes I\right| m\right\rangle. \nonumber
\end{eqnarray} 
Now the LQFI can be obtained as 
\begin{equation}
\mathcal{Q}_A^{2}=1-\lambda^{W}_{max},
\end{equation}
where $\lambda^{W}_{max}$ is the largest eigenvalue of the real symmetric
matrix $W$ with the elements
\begin{equation}\label{matrix1}
[W]_{i j}=\sum_{m \neq n} \frac{2 \lambda_{m} \lambda_{n}}{\lambda_{m}+\lambda_{n}}\left\langle m\left|\sigma_{i} \otimes I\right| n\right\rangle\left\langle n\left|\sigma_{j} \otimes I\right| m\right\rangle.
\end{equation}
\subsection{Local quantum uncertainty}
The uncertainty principle sets a bound on our ability to predict the measurement outcomes of two incompatible observables with arbitrary precision, simultaneously. In general, the uncertainty of measuring a single observable $K$ on a quantum state $\rho$ is defined by variance as
\begin{equation}
Var(\rho,H)=tr\left[ \rho K^{2}\right] -(tr\left[ \rho K\right])^{2}. 
\end{equation}
This uncertainty may include the contributions of classical and quantum nature. In order to determine the quantum part of variance, the concept of skew information is introduced by Wigner and Yanase as \cite{Wigner}
\begin{equation}
I(\rho, K)=-\frac{1}{2} tr[\sqrt{\rho}, K]^{2},
\end{equation}
where $\left[. ,. \right] $ denotes the commutator. It is important to note that unlike variance the Wigner-
Yanase skew information (WYSI) is not affected by classical mixing.  Using the notion of WYSI Girolami et al. introduced a measure for quantum correlations \cite{Girolami}. This measure is called LQU and it is defined by minimizing the WYSI over the local observable  as
 \begin{equation}
 \mathcal{U}(\rho) = \min _{K_{A}} \mathcal{I}\left(\rho, K_{A} \otimes I_{B}\right),
 \end{equation}
 where $K_{A}$ is an observable acting on subsystem $A$. The explicit form of LQU is defined as 
 \begin{equation}
 \mathcal{U}(\rho)=1-max[\lambda_1,\lambda_2,\lambda_3],
 \end{equation}
 where $\lambda_i$'s are eigenvalues of the $3 \times 3$ matrix $M$ with the elements
 \begin{equation}\label{matrix2}
 [M]_{i j} \equiv tr \left\{\sqrt{\rho}\left(\sigma_{i} \otimes I_{B}\right) \sqrt{\rho}\left(\sigma_{j} \otimes I_{B}\right)\right\},
 \end{equation}
where $i,j=1,2,3$ and $\sigma_{i}$'s are Pauli matrices. 
\section{model}\label{Sec3}
Let us consider the bipartite system consist of two-spin anisotropic Heisenberg $XYZ$ chain in the presence of the Dzyaloshinskii–Moriya (DM) interaction. The Hamiltonian of the model is defined as \cite{Radhakrishnan,Zhang}
\begin{equation}
H=J_{x} \sigma_{1}^{x} \sigma_{2}^{x}+J_{y} \sigma_{1}^{y} \sigma_{2}^{y}+J_{z} \sigma_{1}^{z} \sigma_{2}^{z}+\mathbf{D} \cdot\left(\sigma_{1} \times \sigma_{2}\right),
\end{equation}
where $J_k$'s ($k=x,y,z$) are the spin-spin interaction coupling, $\vec{D}$ is the strength of DM interaction and $\sigma_k^{i}$'s ($i=x,y,z$)are  Pauli matrices of $K$-th spin. If the coupling constant $J_i>0$ then the system is antiferromagnetic and if $J_i<0$ then the system is ferromagnetic. In this work we consider the DM interaction in the $z$ direction. The Hamiltonian of $XYZ$ model with DM interaction in the $z$ direction is defined as 
\begin{equation}
H=J_{x} \sigma_{1}^{x} \sigma_{2}^{x}+J_{y} \sigma_{1}^{y} \sigma_{2}^{y}+J_{z} \sigma_{1}^{z} \sigma_{2}^{z}+D_{z}\left(\sigma_{1}^{x} \sigma_{2}^{y}-\sigma_{1}^{y} \sigma_{2}^{x}\right).
\end{equation}
Let us consider $\vert 0 \rangle$ and $\vert 1 \rangle$ as the ground and excited state of a two level particle, respectively. In computational basis $\lbrace \vert 00 \rangle, \vert 01 \rangle, \vert 10 \rangle, \vert 11 \rangle \rbrace$, the Hamiltonian can be written in following matrix form
\begin{equation}H=\left(\begin{array}{cccc}
J_{z} & 0 & 0 & J_{z}-J_{y} \\
0 & -J_{z} & J_{x}+J_{y}+2 i D_{z} & 0 \\
0 & J_{x}+J_{y}-2 i D_{z} & -J_{z} & 0 \\
J_{z}-J_{y} & 0 & 0 & J_{z}
\end{array}\right)
\end{equation}
The spectral analysis leads to following spectrum for the Hamiltonian 
\begin{equation}
E_{1,2}=\pm J_x \mp J_y + J_z, \quad E_{3,4}=-J_z \pm \kappa,
\end{equation}
with
\begin{equation}
\kappa=\sqrt{4 D_{z}^{2}+\left(J_{x}+J_{y}\right)^{2}}.
\end{equation}
The eigenstates of the Hamiltonian are
\begin{equation}
\vert \Phi_{1,2}\rangle=\frac{|00\rangle \pm|11\rangle}{\sqrt{2}} \quad \vert \Phi_{3,4} \rangle=\frac{|01\rangle \pm e^{i \theta}|10\rangle}{\sqrt{2}},
\end{equation}
where 
\begin{equation}
\cos \theta=\frac{J_{x}+J_{y}}{\sqrt{4 D_{z}^{2}+\left(J_{x}+J_{y}\right)^{2}}}.
\end{equation}
Here we want to investigate the temperature dependence of the LQU and LQFI in the two
qubit Heisenberg XYZ model. When the typical solid state system (two-qubit system) is in the thermal equilibrium at temperature $T$, the density operator can be defined as
\begin{equation}
\rho(T)=Z^{-1}e^{-\beta H}=Z^{-1}\sum_{i=1}^{4}e^{- \beta E_i} \vert \Phi_i \rangle \langle \Phi_i \vert,
\end{equation}
where $Z=tr(e^{-\beta H})$ is the partition function of the system, and $\beta = 1/k_B T$ for which $k_B$ is the Boltzmann constant (considered as $k_B=1$ henceforth
for simplicity). Now,the density matrix of the system in thermal equilibrium can be obtained as
\begin{equation}
\rho_{z}(T)=\left(\begin{array}{llll}
r & 0 & 0 & s \\
0 & u & v & 0 \\
0 & v^{*} & u & 0 \\
s & 0 & 0 & r
\end{array}\right)
\end{equation}
where 
\begin{eqnarray}
r&=&\frac{e^{-J_{z} / T}}{Z} \cosh \left(\frac{J_{z}-J_{y}}{2}\right), \nonumber\\
u&=&\frac{e^{J_{z} / T}}{Z} \cosh \left(\frac{\kappa}{T}\right),\nonumber \\
v&=&\frac{e^{J_{z} / T}}{Z \kappa} \sinh \left(\frac{\kappa}{T}\right)\left(2 i D_{z}+J_{x}+J_{y}\right),\nonumber \\
s&=&\frac{e^{-J_{z} / T}}{Z} \sinh \left(\frac{J_{z}-J_{y}}{2}\right).
\end{eqnarray}
The partition function of the system can be written as 
\begin{equation}
Z=2 e^{-J_{z} / T} \cosh \left(\frac{J_{z}-J_{y}}{2}\right)+2 e^{J_{z} / T} \cosh \left(\frac{\kappa}{T}\right).
\end{equation} 
In order to obtain LQFI,  the matrix elements of the matrix $W$  must be specified. According to Eq. \ref{matrix1}, it can be easily shown that the off-diagonal elements  become zero and the diagonal elements are given by
\begin{eqnarray}
W_{11}&=&\frac{4 (r-s) (u-\left| v\right| )}{(u-\left| v\right| )+(r-s)}+\frac{4 (r+s) ( u+\left| v\right|)}{( u+\left| v\right|)+(r+s)}, \nonumber \\
W_{22}&=&\frac{4 (r+s) (u-\left| v\right| )}{(u-\left| v\right| )+(r+s)}+\frac{4 (r-s) (u+\left| v\right| )}{(u+\left| v\right|)+(r-s)}, \nonumber \\
W_{33}&=&\frac{2 \left(u^2-\left| v\right| ^2\right)}{u}+\frac{2 \left(r^2-s^2\right)}{r}.
\end{eqnarray}
So, the LQFI is obtained as
\begin{equation}
\mathcal{Q}_A^{2}=1- \max\lbrace W_{11},W_{22},W_{33}\rbrace. 
\end{equation}
In a similar way, to obtain LQU, the matrix elements of matrix $M$  must be defined. Considering Eq. \ref{matrix2}, it can be easily shown that the off-diagonal elements  equal to zero and the diagonal elements are given by
\begin{eqnarray}
M_{11}&=&2 \left(\sqrt{r-s} \sqrt{u-\left| v\right| }+\sqrt{r+s} \sqrt{ u + \left| v\right|}\right), \nonumber \\
M_{22}&=&2 \left(\sqrt{r+s} \sqrt{u-\left| v\right| }+\sqrt{r-s} \sqrt{u +\left| v\right|}\right), \nonumber \\
M_{33}&=&2 \left(\sqrt{u-\left| v\right| } \sqrt{ u +\left| v\right|}+\sqrt{r-s} \sqrt{r+s}\right).
\end{eqnarray}
So, the LQU is obtained as 
\begin{equation}
\mathcal{U}(\rho)=1-\max \lbrace M_{11}, M_{22}, M_{33} \rbrace.
\end{equation}
\begin{figure}[!]
\centerline{\includegraphics[scale=0.4]{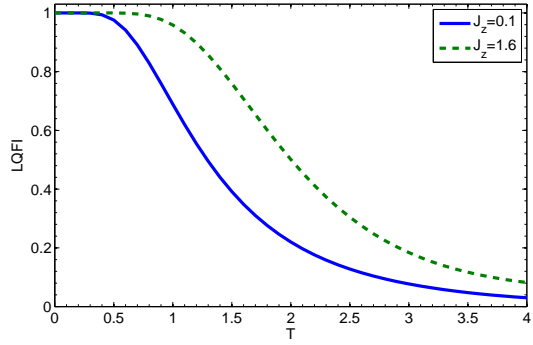}}
\caption{LQFI for the $XYZ$ model with DM interaction oriented along the z-direction as a function of temperature. $J_x=-1$, $J_y=-0.5$ and $D_z=1$. }\label{Figure1}
\end{figure}
In Fig. \ref{Figure1}, LQFI is plotted as a function of temperature  for the XYZ model including the DM interaction
in the z direction. As can be seen due to thermal fluctuations LQFI is decreased by increasing temperature. LQFI  are equal to one at very low temperature and starts to decay only after a thershold temperature.  This happens because thermal fluctuations only affect quantum correlations at temperatures above the characteristic temperature set by the gap energy, which is non-zero for a finite sized systems. We also see that the characteristic temperature at which the LQFI begins to decrease increases with increasing interaction parameter $J_z$.
\begin{figure}[!]
\centerline{\includegraphics[scale=0.7]{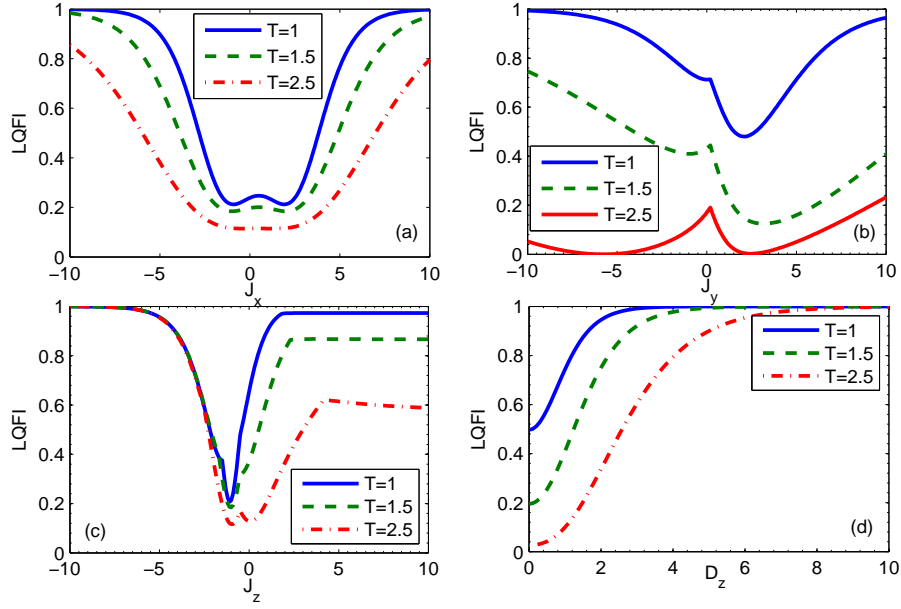}}
\caption{LQFI for the $XYZ$ model with DM interaction oriented along the z-direction. (a)$J_z=-1$, $J_y=-0.5$ and $D_z=1$. (b)$J_x=-1$, $J_z=0.2$ and $D_z=1$. (c)$J_x=-1$, $J_y=-0.5$ and $D_z=1$. (d) $J_x=-1$, $J_y=-1$ and $J_z=0.2$.}\label{Figure2}
\end{figure}
Fig.\ref{Figure2}(a) shows the changes of LQFI in terms of interaction parameter $J_x$. As the value of this parameter increases, the amount of LQFI increases and reaches to constant value one both for the systems with ferromagnetic and anti-ferromagnetic nature. It also can be seen the LQFI is decreased by increasing temperature.
In Fig. \ref{Figure2}(b), the LQFI is plotted as a function of interaction parameter $J_y$. As can be seen the LQFI increases with increasing the value of $J_y$ for both ferromagnetic and anti-ferromagnetic systems. Fig. \ref{Figure2}(c) represents the LQFI interms of interaction parameter $J_z$. As can be seen for ferromagnetic systems LQFI increases and reaches to its maximum value one without dependence on temperature while for anti-ferromagnetic systems LQFI increases and reaches to fixed value. This fixed values varies with the different temperatures. In Fig. \ref{Figure2}(d), the LQFI is plotted as a function of the strength of DM interaction $D_z$. As can be seen LQFI increases with increasing $D_z$ and it always saturates to one. From Figs. \ref{Figure2}(a) and \ref{Figure2}(d) it can be seen that when $J_x$ and $D_z$ increase, the LQFI always saturates to one. This indicates that spin–spin coupling in $x$ and $z$ direction and spin–orbit coupling in $z$ direction can increase  and maintain the value of LQFI of system.
\begin{figure}[!]
\centerline{\includegraphics[scale=0.4]{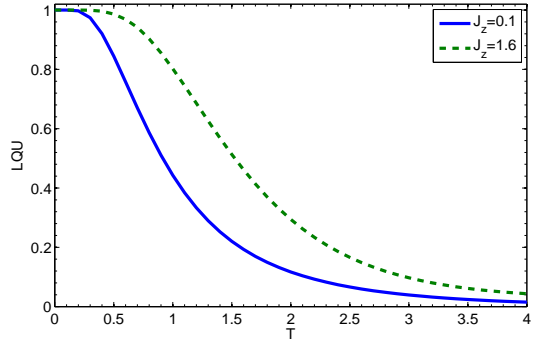}}
\caption{LQU for the $XYZ$ model with DM interaction oriented along the z-direction as a function of temperature. $J_x=-1$, $J_y=-0.5$ and $D_z=1$. }\label{Figure3}
\end{figure}
\begin{figure}[!]
\centerline{\includegraphics[scale=0.7]{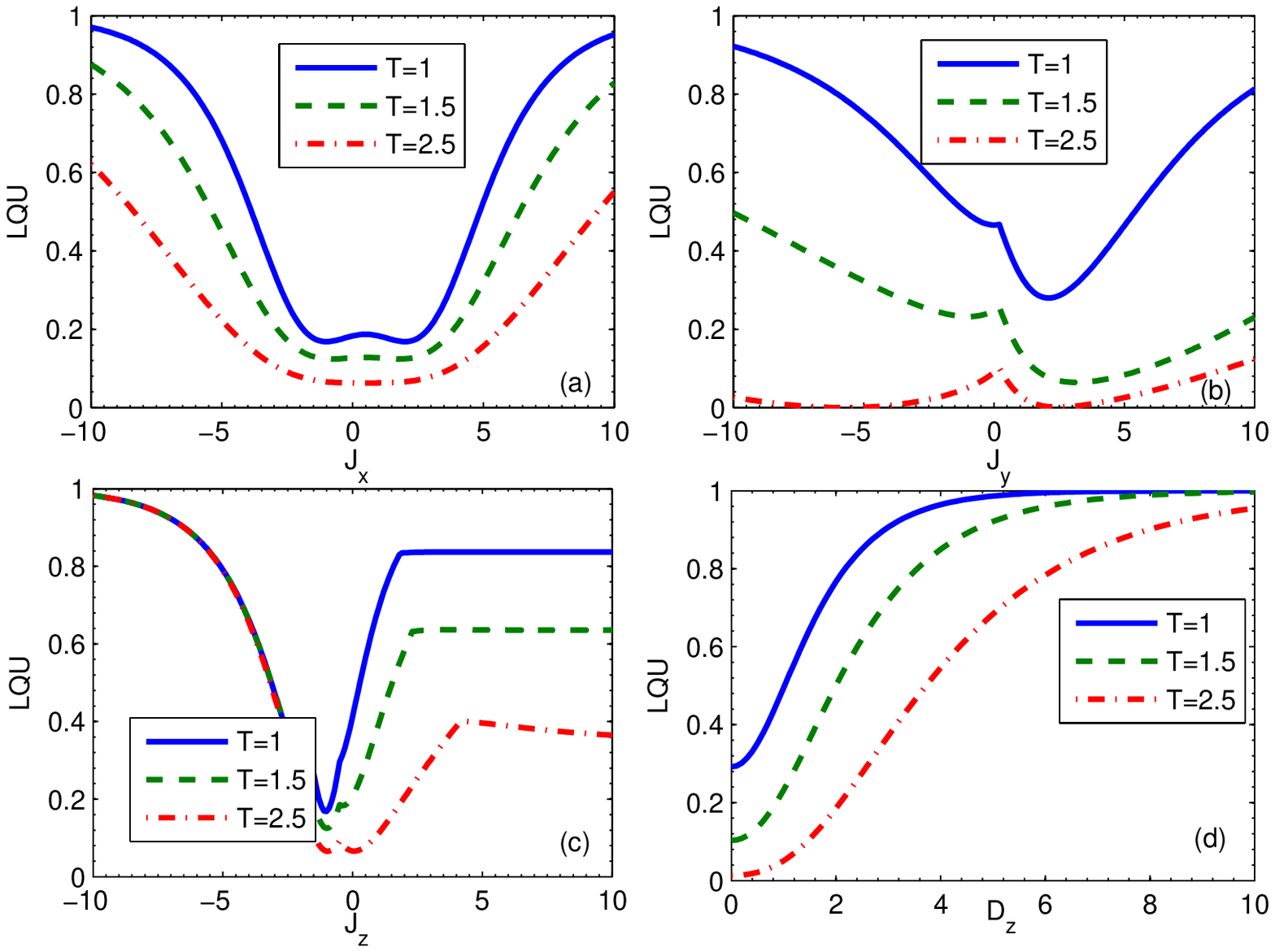}}
\caption{LQU for the $XYZ$ model with DM interaction oriented along the z-direction. (a)$J_z=-1$, $J_y=-0.5$ and $D_z=1$. (b)$J_x=-1$, $J_z=0.2$ and $D_z=1$. (c)$J_x=-1$, $J_y=-0.5$ and $D_z=1$. (d) $J_x=-1$, $J_y=-1$ and $J_z=0.2$. }\label{Figure4}
\end{figure}
In Fig. \ref{Figure3}, LQU is plotted as a function of temperature  for the XYZ model including the DM interaction in the z direction. Figs.\ref{Figure4}(a), \ref{Figure4}(b), \ref{Figure4}(c) and \ref{Figure4}(d) shows the LQU in terms of interaction parameters $J_x$, $J_y$, $J_z$ and $D_z$, respectively.  For LQU the results are quite similar to LQFI. 
\section{conclusion}\label{Sec4}
In this work we have studied the LQFI and LQU in the two-qubit Heisenberg $XYZ$ spin chain model with DM interaction along $z$-direction. We have investigated the effect of temperature on LQFI and LQU for this model. It was shown that due to thermal fluctuations LQFI and LQU are decreased by increasing temperature. They  are equal to one at very low temperature and starts to decay only after a thershold temperature.  This happens because thermal fluctuations only affect quantum correlations at temperatures above the characteristic temperature set by the gap energy, which is non-zero for a finite sized systems. We also see that the characteristic temperature at which the LQFI and LQU begin to decrease increases with increasing interaction parameter $J_z$. We also have shown that LQFI and LQU are increased by increasing interaction parameters. It was shown that when $J_x$ and $D_z$ increase, the LQFI and LQU always saturate to one. This indicates that spin–spin coupling in $x$ and $z$ direction and spin–orbit coupling in $z$ direction can increase  and maintain the value of LQFI and LQU of the system.

\end{document}